\documentclass[12pt]{article}

\ifx\pdfoutput\undefined
\usepackage[dvips,bookmarks]{hyperref}
\else
\usepackage{hyperref}
\fi
\hypersetup{colorlinks=false,bookmarksopen,bookmarksnumbered,citecolor=blue,
   pdfstartview=FitH}

\usepackage{latexsym}

\usepackage{amssymb,amsfonts,amsmath}
\usepackage{graphicx} 
\usepackage{indentfirst}

 \usepackage{bbm}

\parskip=\medskipamount

\def \const {{\rm const}}

\newcommand {\cD}{{\cal D}}

\newcommand {\cH}{{\cal H}}

\newcommand {\cK}{{\cal K}}
\newcommand {\cL}{{\cal L}}

\def\a{\alpha}

\def\b{\beta}
\def\c{\chi}
\def\d{\delta}
\def\e{\epsilon}

\def\g{\gamma}
\def\G{\Gamma}

\def\k{\kappa}
\def\l{\lambda}

\def\q{\theta}
\def\r{\rho}
\def\s{\sigma}

\def\x{\xi}
\def\z{\zeta}

\def\L{\Lambda}

\def\rd{{\rm d}}
\def\ri{{\rm i}}
\def\re{{\rm e}}

\newcommand{\ad}{{\dot{\alpha}}}                           
\newcommand{\bd}{{\dot{\beta}}}                            
\newcommand{\ve}{\varepsilon}                            
\newcommand{\cDB}{{\bar\cD}}                            
\newcommand{\DB}{\bar{D}}

\newcommand{\hf}{\frac12}

\newcommand{\vf}{\varphi}
\newcommand{\be}{\begin{equation}}
\newcommand{\ee}{\end{equation}}
\newcommand{\bea}{\begin{eqnarray}}
\newcommand{\eea}{\end{eqnarray}}
\newcommand{\non}{\nonumber}
\newcommand{\ba}{\begin{array}}
\newcommand{\ea}{\end{array}}

\newcommand{\bm}[1]{\mbox{\boldmath$#1$}}

\def\double #1{#1{\hbox{\kern-2pt $#1$}}}

\newcommand{\gd}{{\dot\g}}

\newcommand{\bsubeq}{\begin{subequations}}
\newcommand{\esubeq}{\end{subequations}}

\newcommand{\qb}{{\bar{\theta}}}

\newcommand{\mub}{{\bar{\mu}}}

\begin{document}
\begin{titlepage}
\begin{flushright}
\end{flushright}

\begin{center}
{\Large \bf 
Free massless higher-superspin superfields on the anti-de Sitter superspace}\footnote{{\it Originally published in}:
Yad.\ Fiz.\  {\bf 57} (1994) 1326; {\it English translation}: Phys.\ Atom.\ Nucl.\  {\bf 57} (1994) 1257. 
The English translation from the Russian original contained some inaccuracies as far as
the scientific terminology is concerned. These inaccuracies have been eliminated 
in the present electronic version.
}
\end{center}

\begin{center}

{\bf
S. M. Kuzenko${}^{2}$
and
A. G. Sibiryakov\footnote{{\it Tomsk State University, Tomsk, 634010 Russia} }
} \\
\vspace{3mm}

\footnotesize{
{\tt Received August 7, 1993}}
~\\
\vspace{2mm}

\vspace{2mm}

\end{center}

\begin{abstract}
\baselineskip=14pt
Free massless higher-superspin superfields on the $N = 1$, $D = 4$ anti-de Sitter superspace are introduced. 
The linearized gauge transformations are postulated. Two families of dually equivalent gauge-invariant action functionals are constructed for massless half-integer-superspin $s+1/2$ $ (s \geq 2)$
and integer-superspin $s$ $(s \geq 1)$ superfields. For $s = 1$, 
one of the formulations for half-integer superspin multiplets reduces to linearized minimal $N = 1$ supergravity 
with a cosmological term, 
while the other is the lifting to the anti-de Sitter superspace of linearized non-minimal $n = -1$ supergravity.
\end{abstract}

\vfill
\end{titlepage}

\newpage
\renewcommand{\thefootnote}{\arabic{footnote}}
\setcounter{footnote}{0}

\tableofcontents{}
\vspace{1cm}
\bigskip\hrule


\allowdisplaybreaks

\section{Introduction}
\setcounter{equation}{0}

Recently, dually equivalent formulations of the theory of arbitrary massless higher-superspin $(s > 3/2)$ 
multiplets\footnote{A massless superspin-$s$ multiplet, 
where $s = 0, 1/2, 1, \cdots,$ describes two ordinary massless particles of spins $s$ and $s + 1/2$ 
and is often denoted by the symbol $(s, s+1/2)$; 
see, e.g., \cite{Thousand} for a review.} were constructed in the $N = 1$, $D = 4$ flat global superspace
 \cite{KPS, KS}. This completed the solution of the long-standing 
probem of finding off-shell (super)field realizations for massless unitary representations of the Poincar\'{e} superalgebra in supersymmetric field theory. 
Earlier, such realizations were known only for lower-superspin $(s \leq 3/2)$ multiplets; the value $s = 3/2$ corresponds to linearized supergravity. 
The formulations presented in \cite{KPS, KS} give an explicit supersymmetric extension of the results obtained in \cite{Fronsdal, FF, Vasiliev}, where 
off-shell field realizations of the massless higher-spin Poincar\'{e} representations were constructed in terms of (non)symmetric tensors and spin-tensors 
on  Minkowski space.

On-shell massless higher-superspin multiplets\footnote{{\it Comment added in 2011:} 
It follows from first principles that the sum of  the actions for free massless spin-$s$ and spin-$(s+1/2)$ fields 
should possess an on-shell supersymmetry.  In this sense, there is no 
problem of constructing on-shell massless higher spin supermultiplets; 
one only needs to work out supersymmetry transformations, which was first done in \cite{Vasiliev, Curtright}. 
The nontrivial problem, however, is  to understand the structure of off-shell 
massless higher spin supermultiplets.}
were described in \cite{Vasiliev, Curtright} on the basis of Lagrangian field formulations 
\cite{Fronsdal, FF, Vasiliev}. An attempt to construct off-shell superfield realizations was made in \cite{BO, BO2}. However, the superfield 
formulation \cite{BO, BO2} does not correspond to specific features of genuinely massless higher-super-spin multiplets. Supermultiplets 
arising within the scheme proposed in \cite{BO, BO2} are too reducible, 
and additional constraints should be imposed 
for eliminating extra degrees of freedom. In \cite{BO2}, such constraints were found only for the gravitino multiplet (superspin $s = 1$); however, 
correct off-shell realizations of this multiplet were constructed much earlier \cite{OS, FV, deWvanH, GS}. 
Moreover, it can be shown that the 
supefield formulation \cite{BO, BO2} does not admit a generalization to 
the anti-de Sitter (AdS) superspace \cite{Keck, Zumino, IS}.

The AdS superspace is a special solution of the equations of motion of (old minimal) $N = 1$, $D = 4$ superfield 
supergravity with a cosmological term (see \cite{Thousand, Ideas} for reviews). 
In the AdS superspace, a surface parametrized only by even variables coincides with the ordinary AdS space. 
In  the family of curved $N = 1$, $D = 4$ supergeometries, the flat superspace and the AdS superspace 
are the only ones (up to transformations of equivalence) that possess a constant supertorsion 
and maximal  symmetry supergroup. Symmetries of the AdS 
superspace are generated by transformations that belong to the  superalgebra $\rm{osp}(1, 4)$,
whose even part coincides with the AdS algebra $\rm{sp}(4) \simeq \rm{so}(3,2)$.

In this work, we generalize the results obtained in \cite{KPS, KS} to the case of the AdS superspace. 
We introduce massless higher-superspin superfields, 
postulate their inearized gauge transformations, and construct free gauge-invariant action functionals. 
For each free massless higher-superspin $(s > 3/2)$ multiplet, 
we propose two classically equivalent Lagrangian formulations, which are related to each other 
by a duality transformation. In the limit of vanishing curvature of the AdS superspace, these formulations  
reduce to the theories in the flat superspace which were constructed in \cite{KPS, KS}.

${}$From the viewpoint of group theory, the proposed theories 
realize massless representations of the superalgebra  $\rm{osp}(1, 4)$ on the mass shell. 
It is well known \cite{Heidenreich, NdeWFvanN} that the irreducible unitary massless representations of 
the superalgebra $\rm{osp}(1, 4)$ 
 are classified by the superspin $s = 0, 1/2, 1, \cdots .$ 
 With respect to the AdS algebra $\rm{sp}(4) \simeq \rm{so}(3, 2)$, 
the superspin-$s$ $(s \neq 0)$ representation decomposes 
into the  sum $D(s+1, s) \oplus D(s + 3/2 , s + 1/2)$ of two irreducible massless 
spin-$s$ and spin-$(s + 1/2)$ representations of $\rm{sp}(4)$, where $D(E_0, s)$ 
is the representation of $\rm{sp}(4)$ with spin $s$ and 
mass $m^2 = E_0(E_0 - 3) - (s + 1)(s - 2)$ ($E_0$ is the lowest energy) \cite{NdeWFvanN, Fronsdal2, Evans}. 
In the limit of vanishing curvature 
of the AdS space, the massless spin-$s$ representation $D(s + 1, s)$ can be reduced 
to the massless helicity-$s$ (or $-s$) representation of the Poincar\'{e} algebra \cite{AFFS}. 
The theory of free superfields in AdS superspace is said  to describe a massless superspin-$s$ multiplet 
if the corresponding representation of $\rm{osp}(1, 4)$, acting on the mass shell, is given by the sum of two (conjugate) massless superspin-$s$ representations and, 
in the limit of vanishing curvature, is reduced to the tensor sum of the four 
Poincar\'{e} representations with the helicities $\pm s$ and $\pm (s + 1/2)$.

In accordance with the terminology of the Batalin-Vilkovisky field-antifield quantization \cite{BV}, 
the gauge superfield theories proposed in our work are reducible. 
The stage of reducibility can be finite or infinite depending on the type of the formulation and on the superspin. 
In this sense, the higher-superspin models of an infinite stage of reducibility are similar to the Green-Schwarz formulation of superstring theory \cite{GS2}. 
A remarkable feature of the proposed models on the AdS superspace is that 
there is always a covariant reparametrization of generators 
of gauge transformations that converts an infinite stage of reducibility to a finite one and vice versa. 
For this reason, our models can be used 
as nontrivial playgrounds for developing the methods of covariant quantization.

Recently, considerable advances have been made, 
especially in the papers by Vasiliev and Fradkin \cite{VF, FV2, Vasiliev2, Vasiliev3, Vasiliev4}, 
toward the solution of the famous problem of higher spins. Among other things, it was shown in \cite{VF} 
that gravitational interaction of higher-spin 
fields can be introduced in a consistent manner at least in the lowest order; however, 
it turns out to be non-analytic in the cosmological constant. 
Therefore, the flat space-time cannot be used as a classical vacuum in consistent theories of interacting 
massless higher-spin fields. Only the 
expansion around the AdS background is well defined for such theories. We believe that the results of 
our work provide necessary input data 
for developing the superfield approach to the problem of higher spins.

Our paper is organized as follows. In Section 2, we present basic information on the geometry 
of the AdS superspace and introduce transverse 
and longitudinal linear tensor superfields, 
which give the realization of irreducible off-shell superfield representations of the Grassmann
envelope of the $\rm{osp}(1, 4)$ superalgebra \cite{IS}. In Section 3, we construct the Lagrangian 
formulations for free massless 
half-integer-superspin $(s > 3/2)$ multiplets. A similar analysis for integer-superspin $(s \geq 1)$ multiplets 
is carried out in Section 4. 
The component analysis is given in Section 5. Here, we show that the action functional 
for the massless half-integer-superspin $(s + 1/2, s \geq 2)$ 
multiplet in the transverse formulation reduces in components to the sum of actions for two massless fields 
with spin $s + 1$ \cite{Fronsdal3} and $s + 1/2$ \cite{FF2} on the AdS space. 
The characteristic features of the supergravity multiplet (the superspin $3/2$) are discussed in Section 6.

For the most part, our notation coincides with that used by Wess and Bagger \cite{WB}; some specific points and the most useful identities 
and relations are given in the appendix.

\section{The AdS superspace}

The AdS superspace \cite{Keck, Zumino, IS} is  a special curved $N = 1$, $D = 4$ superspace. 
Assuming that it is parametrized by  local coordinates 
$z^M = (x^m, \q^\mu, \bar{\q}_{\dot{\mu}})$, the geometry of this space is specified by 
the supervielbein $E_A = E_A{}^M(z) \partial_M$ and the Lorentz superfield connection
\be 
\Omega_A = \hf \Omega_A{}^{bc} M_{bc} = \Omega_A{}^{\b\g} M_{\b\g} 
+ \bar{\Omega}_A{}^{\bd\gd} \bar{M}_{\bd\gd} \ , \non
\ee
such that the corresponding covariant derivatives
\be \cD_A = (\cD_a , \cD_\a , \cDB^\ad) = E_A + \Omega_A 
\label{2.1}
\ee
satisfy the algebra
\bea
\{ \cD_\a, \cDB_\ad \} &=& - 2 \ri \cD_{\a\ad} \ , \non\\
\{ \cD_\a , \cD_\b \} &=& - 4 \mub M_{\a\b} \ , \quad \{ \cDB_\ad , \cDB_\bd \} = 4 \mu \bar{M}_{\ad\bd} \ , \non\\
\left[ \cD_\a , \cD_{\b \bd} \right] &=& \ri \mub \ve_{\a\b} \cDB_\bd \ , \quad [\cDB_\ad , \cDB_{\b\bd}] 
= - \ri \mu \ve_{\ad\bd} \cD_\b \ , \non\\
\left[ \cD_{\a\ad} , \cD_{\b\bd} \right] &=& - 2 \mub \mu (\ve_{\a\b} \bar{M}_{\ad\bd} + \ve_{\ad\bd} M_{\a\b}) \ , 
\label{2.2}
\eea
where $\mu$ is a nonzero constant having the dimension of mass (our notation is described in the appendix).
 In our case, the components 
$(R, G_a, W_{\a\b\g})$ of the supertorsion, which characterize geometry of the curved superspace \cite{WB}, 
have the form
\be 2 R = \mu = \const \ , \quad G_a = W_{\a\b\g = 0} \ . \label{2.3}
\ee
Algebra \eqref{2.2} is invariant under arbitrary superspace coordinate transformations and local Lorentz transformations with parameters of the form
\be \cK = K^M(z) \partial_M + K^{\a\b} (z)M_{\a\b} + \bar{K}^{\ad\bd}(z) \bar{M}_{\ad\bd} = \bar{\cK} \ , 
\label{2.4}
\ee
acting on covariant derivatives and on arbitrary tensor superfield $V_{\a(r)\ad(t)}$ of the Lorentz type $(r/2, t/2)$ 
on the AdS superspace according to the 
law
\be \d \cD_A = \left[ \cK , \cD_A \right] \ , \quad \d V_{\a(r)\ad(t)} = \cK V_{\a(r) \ad(t)} \label{2.5} \ .
\ee

It is well known (see, e.g., \cite{Ideas}) that, in the old minimal $N = 1$, $D = 4$ supergravity, 
the gauge arbitrariness specified by parameters \eqref{2.4} makes 
it possible to express all geometric objects $E_A{}^M$ and $\Omega_A{}^{bc}$ in terms of the Ogievetsky-Sokatchov gravitational superfield $\cH^m(z) = \bar{\cH}^m$,
the Siegel-Gates chiral compensator $\vf(z) = \vf(x^m + i \cH^m, \q^\mu)$ and its conjugate $\bar{\vf}(z)$ 
(the corresponding expressions are given in the appendix). 
A specific feature of the AdS superspace is 
that the remaining gauge freedom can be partially used for imposing the Wess-Zumino gauge
\bea
\cH^m(x, \q, \bar{\q}) &=& e_a{}^m(x) \left[ \q \s^a \bar{\q} + \frac{1}{4} \q^2 \bar{\q}^2 \ve^{abcd} w_{bcd}(x) \right] \ , \non\\
\vf(x, \theta) &=& e^{-1/3}(x) \left[ 1 + \q^2 \mub \right] \ , \qquad e \equiv \det{e_a{}^m} \label{2.6} \ ,
\eea
where the space-time vierbein $e_a{}^m$ and the Lorentz connection $w_{bcd} = - w_{bdc}$ 
specify the geometry of the AdS superspace:
\bea
\nabla_a = e_a{}^m(x) \partial_m &+& \hf w_a{}^{bc}(x) M_{bc} \ , \non\\
\left[ \nabla_a , \nabla_b \right] &=& - |\mu|^2 M_{ab} \ .
\eea
The covariant derivatives $\cD$ and $\nabla$ are related to each other by a very simple equation, namely, introducing the projection
\be V_{\a(r)\ad(t)}| \equiv V_{\a(r)\ad(t)}(x, \q = 0, \bar{\q} = 0)
\ee
of the tensor superfield on its zero component with respect to $\q$ and $\bar{\q}$, we easily obtain
\be (\cD_a V_{\b(r) \bd(t)})| = \nabla_a V_{\b(r) \bd(t)} | . \label{2.9}
\ee
The remaining transformations \eqref{2.4}, \eqref{2.5} surviving in the gauge \eqref{2.6} are generated by the parameters of two types, namely, 
\begin{subequations}
\bea
K^m &=& \hf b^m(x + \ri \cH) + \hf b^m(x - i \cH) \ , \quad b^m = \bar{b}^m \ , \non\\
K^\mu &=& K^{\a\b} = 0 \ ; \\
K^m &=& 0 \ , K^\mu = \d^\mu_\a \L^\a{}_\b(x + \ri \cH) \d^\b_\mu \q^\mu \ , \quad \L^{\a\b} = \L^{\b\a} \ , \non\\
K^{\a\b} &=& \hat{E}^{(\a} K^{\b)} \ , \quad K^\a = \d^\a_\mu K^\mu \ ,
\eea
\end{subequations}
where the operator $\hat{E}_\a$ is defined in \eqref{A.11}. 
They correspond to space-time coordinate transformations $(b^m)$ and local Lorentz 
transformations $(\L^{\a\b})$. It is seen that, in the gauge \eqref{2.6}, 
the AdS space is identified with the surface $\q = \bar{\q} = 0$ in the AdS superspace.

Another convenient gauge fixing is based on the fact that the AdS (super)space is (super)conformally flat. 
As a consequence, gauge arbitrariness \eqref{2.4}, 
\eqref{2.5} can be used in such a way as to reduce covariant derivatives \eqref{2.1} to the  form
\bea
\cD_\a &=& F D_\a - 2 (D^\b F) M_{\a\b} \ , \quad F = \vf^{1/2} \bar{\vf}^{-1} \ , \non\\
\cDB_\ad &=& \bar{F} \DB_\ad - 2 (\DB^\bd \bar{F}) \bar M_{\ad\bd} \ . \label{2.11}
\eea
Here, $D_\a$ and $\DB_\ad$ are the spinor covariant derivatives of the flat global superspace, 
and the chiral compensator has the form
\bea \vf(z) &=& \re^{\ri \cH_0} \left( 1 - \frac{1}{4} |\mu|^2 x^2- \mub \q^2 \right)^{-1} \ , \qquad
\cH_0 = \q \s^a \bar{\q} \d_a^m \partial_m \ . \label{2.12}
\eea
A detailed account of the formulated statements can be found in the monograph \cite{Ideas}.

Symmetries of the AdS superspace are associated with the operators $\cK_0$ having the form \eqref{2.4} 
and satisfying the condition
\be [ \cK_0, \cD_A] = 0 \ . 
\ee
The set of all such operators is isomorphic to a real Grassmann envelope of the superalgebra $\rm osp(1, 4)$.  
In the gauge specified by 
\eqref{2.11} and \eqref{2.12}, we have
\be i \cK_0 = \hf \L^{\a\ad} {\bm P}_{\a\ad} + \L^{\a\b} {\bm J}_{\a\b} + \bar{\L}^{\ad\bd} \bar{\bm J}_{\ad\bd} + \L^\a {\bm Q}_\a + \bar{\L}_\ad \bar{\bm Q}^\ad \ ,
\ee
where the parameters $\L$'s are (anti)commuting numbers, and the explicit form of the generators is given by the expressions \eqref{A.14}. The generators specify 
the action of $\rm osp(1, 4)$ on the tensor superfields in the AdS superspace.

Complex tensor superfields $\G_{\a(r) \ad(t)}$ and $G_{\a(r) \ad(t)}$ are referred to as 
transverse and longitudinal linear superfields, respectively, if the constraints
\begin{subequations} \label{2.15}
\bea
\cDB^\bd \G_{\a(r) \bd \ad(t - 1)} &=& 0 \ , \quad t \neq 0 \ ; \label{2.15a}\\
\cDB_{(\bd} G_{\a(r)\ad(t) )} &=& 0 \label{2.15b}
\eea
\end{subequations}
are satisfied. At $t = 0$, the constraint \eqref{2.15b} reduces to the condition of covariant chirality
\be \cDB_\bd G_{\a(r)} = 0 \ .
\ee
In contrast to the flat superspace, the condition of covariant chirality is inconsistent for $t \neq 0$ because we have
\be \cDB_\bd \Phi_{\a(r) \ad(t)} = 0 \rightarrow \Phi_{\a(r) \ad(t)} = 0 \ , \quad t \neq 0 \non
\ee
as a consequence of algebra \eqref{2.2}. The off-shell contraints \eqref{2.15} are the ``strongest'' in the AdS superspace 
\cite{IS}. The relations \eqref{2.15} lead to the linearity conditions
\begin{subequations}
\bea
(\cDB^2 - 2 (t + 2) \mu) \G_{\a(r) \ad(t)} = 0 \label{2.17a} \ , \\ 
(\cDB^2 + 2 t \mu ) G_{\a(r) \ad(t)} = 0 \ . \label{2.17b}
\eea
\end{subequations}

The transverse and longitudinal linear superfields, which were used in \cite{KPS, KS} for 
constructing theories of free massless higher-superspin 
superfields in the flat superspace,  are actually well defined on an arbitrary supergravity background.

The constraints \eqref{2.15} can be solved in terms of complex 
unconstrained tensor superfields $\x_{\a(r) \ad(t+1)}$ and $\z_{\a(r) \ad(t-1)}$ according to the rules 
\begin{subequations}
\bea
\G_{\a(r) \ad(t)} &=& \cDB^\bd \x_{\a(r) \bd \ad(t)} \ , \label{2.18a} \\ 
G_{\a(r)\bd\ad(t-1)} &=& \cDB_{(\bd} \z_{\a(r) \ad(t-1))} \ . \label{2.18b}
\eea
\end{subequations}
There is a natural arbitrariness in the choice of the potentials $\x$ and $\z$, namely,
\begin{subequations}
\bea
\d \x_{\a(r)\ad(t+1)} &=& \g_{\a(r) \ad(t+1)} \ , \\
\d \z_{\a(r) \ad(t-1)} &=& g_{\a(r) \ad(t - 1)} \ .
\eea
\end{subequations}
Here, $\g$ is a transverse linear superfield, and $g$ is a longitudinal linear superfield. Thus, in accordance with the terminology of gauge theories with linearly 
dependent generators \cite{BV}, any Lagrangian theory described by a transverse  linear superfield $\G_{\a(r)\ad(t)}$ (a longitudinal linear superfield 
$G_{\a(r)\ad(t)}$) can be considered as the theory of a general superfield $\x_{\a(r)\ad(t+1)}$ 
($\z_{\a(r)\ad(t-1)}$) with an additional gauge invariance of an infinite 
stage of reducibility (of the $(t-1)$th stage of reducibility).

It is important to note that the existence of the representations \eqref{2.18a} and \eqref{2.18b} 
is a corollary to the identity
\bea
V_{\a(r) \bd \ad(t-1)} = &-& \frac{1}{2 \mu (t+2)} \cDB^\gd \cDB_{(\gd} V_{\a(r) \bd \ad(t-1))} \non\\
&-& \frac{1}{2 \mu (t+1)} \cDB_{(\bd} \cDB^{|\gd|} V_{\a(r) \ad(t-1) ) \gd} \ , \non
\eea
which holds for an arbitrary tensor superfield $V_{\a(r)\ad(t)}$ on the AdS superspace, provided that $t \neq 0$. Therefore, a general superfield $V_{\a(r) \ad(t)}$ admits 
the decomposition of the form
\be V_{\a(r) \ad(t)} = \G_{\a(r) \ad(t)} + G_{\a(r) \ad(t)} \ , \non
\ee
where $\G$ and $G$ are given by the relations \eqref{2.18a} and \eqref{2.18b}.

\section{Theories of free massless superfields with higher half-integer superspins}

To obtain a realization of the free massless superspin-$(s + 1/2)$  multiplet  $(s = 2, 3, \cdots)$
in the Lagrangian approach, we introduce two sets of tensor boson superfields 
on the AdS superspace, namely
\begin{subequations} \label{3.1}
\bea
v^{\perp}_{(s+1/2)} &=& \left\{ H_{\a(s)\ad(s)}(z), \, \G_{\a(s-1)\ad(s-1)}(z),\, \bar{\G}_{\a(s-1) \ad(s-1)} (z) \right\} \ ;
 \label{3.1a} \\
v^{||}_{(s + 1/2)} &=& \left\{ H_{\a(s) \ad(s)} (z) , \, G_{\a(s-1) \ad(s-1)}(z), \, \bar{G}_{\a(s-1) \ad(s-1)}(z) \right\} \ ,
 \label{3.1b}
\eea
\end{subequations}
which will play the role of dynamical variables in two different, 
but on-shell equivalent, locally supersymmetric theories.
 In both cases, $H_{\a(s)\ad(s)}$ is a  real unconstrained superfield of the Lorentz type $(s/2, s/2)$. 
However, the complex superfields $\G_{\a(s-1)\ad(s-1)}$ and $G_{\a(s-1) \ad(s-1)}$ satisfy the transverse linear 
and longitudinal linear constraints, 
respectively.

We postulate linearized gauge transformations of the superfields $H$, $\G$ and $G$ in the form
\begin{subequations} \label{3.2}
\bea \d H_{\b \a(s-1) \bd \ad(s-1)} = &&\cDB_{(\bd} L_{\b \a(s-1) \ad(s-1) )} \non\\
&&- \cD_{(\b} \bar{L}_{\a(s-1))\bd \ad(s-1)} \ , \label{3.2a} \\
\d \G_{\a(s-1)\ad(s-1)} = &&-\frac{s}{2 (s+1)} \cDB^\gd \cD^\g \cD_{(\g} \bar{L}_{\a(s-1))\gd\ad(s-1)} \ , \label{3.2b} \\
\d G_{\a(s-1)\bd\ad(s-2)} = && - \hf \cDB_{(\bd} \cDB^{|\gd|} \cD^\g L_{\g\a(s-1)\ad(s-2)) \gd} \non\\
&&+ \ri (s-1) \cDB_{(\bd} \cD^{\g |\gd|} L_{\g \a(s-1) \ad(s-2) ) \gd} \ . \label{3.2c}
\eea
\end{subequations}
Here, $L_{\a(s)\ad(s-1)}$ is an arbitrary fermionic superfield of the Lorentz type $(3/2, (s-1)/2)$. 
Symmetrization in \eqref{3.2c} is extended only to the subscripts $\bd$ and $\ad(s-2)$. 
In the flat-space limit $(\mu \rightarrow 0)$, expressions \eqref{3.2} reduce to the the gauge transformations of massless higher-superspin superfields on the flat superspace 
\cite{KPS}.

In addition to the requirement that the flat-space limit should be correct, we were guided by the following considerations in choosing the transformation laws in the form \eqref{3.2}. 
First of all, expression \eqref{3.2a} generalizes the linearized gauge transformation law of
 the gravitational superfield $H_{\a\ad}$ in supergravity (superspin $3/2$) \cite{Thousand, Ideas}. 
Now, we notice that the right-hand side of \eqref{3.2a} is the sum of the longitudinal linear superfield
\be \cDB_{(\bd} L_{\a(s) \ad(s-1))} \label{3.3}
\ee
and its conjugate. The second guiding principle in determining the structure of gauge transformations 
is that variations of the compensating superfields $\G$ and $G$ 
must not involve any fields other than the superfield \eqref{3.3} and its conjugate. 
This requirement, supplemented with the condition of longitudinal linearity, fixes 
(up to a constant factor) the transformation \eqref{3.2c} because this expression can be rewritten as
\bea \d G_{\a(s-1) \ad(s-1)} = \frac{s}{2(s+1)} \cD^\g \cDB^\gd \cDB_{(\gd} L_{\g\a(s-1)\ad(s-1))} + i s \cD^{\g\gd} \cDB_{(\gd} L_{\g \a(s-1) \ad(s-1))} \ . \non
\eea
We can generalize the transformation \eqref{3.2b} by making the substitution $\G \rightarrow \widetilde{\G}$, 
where
\be 
\widetilde{\G}_{\a(s-1)\ad(s-1)} = \G_{\a(s-1)\ad(s-1)} + c \cDB^\bd \cD^\b H_{\b \a(t-1) \bd \ad(s-1)}  \non
\ee
and $c$ is a constant. However, it is the choice of the expression \eqref{3.2b} that leads to the simplest 
structure of the gauge-invariant action functional (see below). The stage 
of reducibility of the gauge transformation \eqref{3.2} is $s$.

Apparently, concepts underlying the procedure of fixing the structure of gauge transformations 
should be sought for at the nonlinear level. For example, if only the requirement 
that the flat-space limit should be correct is used as the guiding principle, 
the transformation \eqref{3.2b} can be generalized by including terms proportional to the curvature of the 
AdS superspace. In this case, we have
\bea
\d \G_{\a(s-1)\ad(s-1)} = - \frac{s}{2 (s+1)} \cDB^\gd \cD^\g \cD_{(\g} \bar{L}_{\a(s-1)) \gd \ad(s-1)} 
+ (\e_1 \mu + \e_2 \mub) \cDB^\gd \bar{L}_{\a(s-1)\gd\ad(s-1)} \ , \non
\eea
where $\e_1$ and $\e_2$ are dimensionless constraints. 
However, it can easily be shown that the gauge-invariant action of the superfields \eqref{3.1a} 
exists only for  $\e_1 = \e_2 = 0$.

Let us now proceed to the construction of gauge-invariant action functionals. 
We fix the mass dimensions of the superfields \eqref{3.1a} according to the rule
\be 
[H_{\a(s)\ad(s)}] = 0 \ , \qquad [\G_{\a(s-1)\ad(s-1)}] = 1 \non
\ee
and seek a quadratic local functional of $H$, $\G$, and $\bar{\G}$ that is invariant under 
the transformations \eqref{3.2a} and \eqref{3.2b}. Such a functional proves to be unique 
up to a constant factor and is given by the expression
\bea
S^{\perp}_{(s+1/2)} &= &\Big( - \frac{1}{2} \Big)^s \int \rd^8z \,E^{-1} \Big\{ 
\frac{1}{8} H^{\a(s) \ad(s)} \cD^\b (\cDB^2 - 4 \mu)
\cD_\b H_{\a(s)\ad(s)} 
 \non\\
&&+ H^{\b \a(s-1) \bd \ad(s-1)} (\cD_\b \cDB_\bd \G_{\a(s-1) \ad(s-1)}
- \cDB_\bd \cD_\b \bar{\G}_{\a(s-1)\ad(s-1)} ) 
 \non\\
&&+ \frac{s^2}{2} \bar{\mu} \mu H \cdot H 
+ 2 \bar{\G} \cdot \G + \frac{s+1}{s} (\G \cdot \G + \bar{\G} \cdot \bar{\G}) \Big\} \ . \label{3.4}
\eea
Here, $\rd^8z = \rd^4x\rd^2\q\rd^2\qb$ is the superspace integration measure, 
$E = {\rm Ber}(E_A{}^M)$, and the dots in the combinations $H \cdot H$ and $\bar{\G} \cdot \G$ 
denote the operation of complete contraction of the indices defined by eq. \eqref{A.6}. 
The first term on the right-hand side of \eqref{3.4} is real-valued as a 
consequence of \eqref{A.11}. The presented action is manifestly invariant under superspace coordinate transformations and local Lorentz transformations.

In the limit of vanishing curvature $(\mu \rightarrow 0)$, expression \eqref{3.4}  
reduces to the action functional of the free massless superspin-$(s + 1/2)$ multiplet in the 
flat superspace in the transverse formulation \cite{KPS}. It will be shown in Section 5 that, 
upon elimination of the  auxiliary fields, the theory with the action functional \eqref{3.4} 
considered at the component level describes free massless fields of spins $s + 1/2$ and $s + 1$ 
in the AdS space. Hence, we obtained the Lagrangian realization of the free 
massless superspin-$(s+1/2)$ multiplet on the AdS superspace.

The above theory can be reformulated in terms of the superfields \eqref{3.1b} by using a special duality transformation. Let us introduce the auxiliary model characterized 
by the action functional
\bea
S[H, G, V] & = &\Big( - \frac{1}{2} \Big)^s \int \rd^8z \,E^{-1} \Big\{ \frac{1}{8} H^{\a(s) \ad(s)} \cD^\b 
(\cDB^2 - 4 \mu) \cD_\b H_{\a(s)\ad(s)} \non \\
&& 
+ \frac{s^2}{2} \bar{\mu} \mu H \cdot H 
+ \Big(H^{\b\a(s-1)\bd\ad(s-1)} \cD_\b \cDB_\bd V_{\a(s-1)\ad(s-1)} \non\\
&&- \frac{2}{3} G \cdot V + \bar{V} \cdot V + \frac{s+1}{s} V \cdot V + \rm{h.c.}   \Big) \Big\} \ , \label{3.5}
\eea
where $V_{\a(s-1)\ad(s-1)}$ is  a complex unconstrained superfield of the Lorentz type $((s-1)/2, (s-1)/2)$. 
Under
the equations of motion for $G$ and $\bar{G}$, 
the action functional \eqref{3.5} reduces to \eqref{3.4}. Indeed, in varying \eqref{3.5} with respect to $G$, 
we must take into account the constraint \eqref{2.15b} or, what is the same, 
use the representation \eqref{2.18b}. This yields
\bea
\d_G S[H, G, V] &=& \int \rd^8 z \,E^{-1} \d G \cdot V \non\\
&=& \int \rd^8 z E^{-1} \d \z^{\a(s-1) \ad(s-2)} \cDB^\bd V_{\a(s-1)\bd \ad(s-2)} \ . \non
\eea
We see that under the equation of motion for $G$,
the superfield $V_{\a(s-1) \ad(s-1)}$ becomes transverse linear and can be identified with 
$\G_{\a(s-1)\ad(s-1)}$. As a result, the action functional \eqref{3.5} goes over to \eqref{3.4}. 
On the other hand, we can eliminate the superfield $V$ from $S[H, G, V]$ using the 
equation of motion
\be
s \bar{V}_{\a(s-1) \ad(s-1)} + (s+1) V_{\a(s-1) \ad(s-1)} - \frac{s}{2} \cDB^\bd \cD^\b H_{\b \a(s-1) \bd \ad(s-1)} - G_{\a(s-1) \ad(s-1)} = 0 \non
\ee
obtained by varying \eqref{3.5} with respect to $V$. This procedure leads to the action functional
\bea
S^{||}_{(s+1/2)} &=& \Big( - \frac{1}{2} \Big)^s \int \rd^8 z \,E^{-1} 
\Big\{ \frac{1}{8} H^{\a(s) \ad(s)} \cD^\b(\cDB^2 - 4 \mu) 
\cD_\b H_{\a(s) \ad(s)} 
\non \\
&&- \frac{1}{8} \frac{s}{2s + 1} [\cD_\b , \cDB_\bd] H^{\b \a(s-1) \bd \ad(s-1)} 
[\cD^\g, \cDB^\gd] H_{\g \a(s-1) \gd \ad(s-1)} \non \\
&&+ \frac{s}{2} \cD^{\b \bd} H_{\b \a(s-1) \bd \ad(s-1)} 
\cD_{\g\gd} H^{\g \a(s-1) \gd \ad(s-1)} + \frac{s^2}{2} \mub \mu H \cdot H
\non \\
&&+ \frac{2 i s}{2s+1} \cD_{\b\bd} H^{\b \a(s-1) \bd \ad(s-1)}
(G_{\a(s-1) \ad(s-1)} - \bar{G}_{\a(s-1) \ad(s-1)})  \non\\
&&+ \frac{2}{2 s + 1} \bar{G} \cdot G + \frac{s+1}{s(2 s + 1)} (G \cdot G + \bar{G} \cdot \bar{G}) \Big\} \ . \label{3.6}
\eea
It is easy to verify that the functional \eqref{3.5} is invariant under the transformations \eqref{3.2a} and \eqref{3.2c} supplemented with the variation $\d V_{\a(s-1) \ad(s-1)}$  of the form \eqref{3.2b}. 
If follows that action \eqref{3.6} is invariant under the transformations \eqref{3.2a} and \eqref{3.2c}. 
It is seen from the above analysis that theories 
characterized by the actions \eqref{3.4} and \eqref{3.6} generate the same dynamics.

As in the case of the flat superspace \cite{KPS}, the two formulations of the theory of 
the massless superspin-$(s+1/2)$ multiplet that are specified by the actions \eqref{3.4} and 
\eqref{3.6} can be referred to as transverse linear and longitudinal linear formulations, respectively.

Without going into details, we note that in both formulations the physical sector of the theory is described by two gauge-invariant field strengths, namely, by the chiral tensor superfield
\bea
W_{(\a_1 \cdots \a_{2s + 1})} &=& (\cDB^2 - 4 \mu) \cD_{(\a_1}{}^{\bd_1} \cdots \cD_{(\a_s}{}^{\bd_s} \cD_{\a_{s+1}} H_{\a_{s+2} \cdots \a_{2s+1}) \bd_1 \cdots \bd_s} \ , \non\\
\cDB_\bd W_{\a(2s+1)} = 0 \label{3.7}
\eea
and its conjugate $\bar{W}_{\ad(2s+1)}$. In the flat-space limit, the quantity $W_{\a(2s+1)}$ becomes an on-shell massless superfield with superhelicity 
$s+1/2$ (helicities $s+1/2$ and $s+1$ for the component fields $W_{\a(2s+1)}$) \cite{Thousand, Ideas}.

\section{Theories of free massless superfields with higher integer superspins}

A free massless multiplet of integer superspin $s$ $(s = 1,2, \cdots)$  
can be realized in the AdS superspace by using two different sets of tensor superfields having the form
\begin{subequations}  \label{4.1}
\bea
v^\perp_{(s)} &=& \left\{ H_{\a(s-1) \ad(s-1)}(z), \G_{\a(s)\ad(s)}(z), \bar{\G}_{\a(s)\ad(s)}(z) \right\} \ , \label{4.1a} \\
v^{||}_{(s)} &=& \left\{ H_{\a(s-1) \ad(s-1)}(z), G_{\a(s)\ad(s)}(z), \bar{G}_{\a(s)\ad(s)}(z) \right\} \ . \label{4.1b}
\eea
\end{subequations}
Here, $H_{\a(s-1)  \ad(s-1)}$ is a general real superfield, $\G_{\a(s) \ad(s)}$ is a transverse linear superfield, and $G_{\a(s)\ad(s)}$ is a longitudinal linear superfield. 
Comparison with \eqref{3.1} shows that, for $s \geq 2$, the kinematic difference 
between the cases of the half-integer $(s+1/2)$ and the integer $(s)$ superspins 
amounts to the 
inversion of the tensor structure between $H$ and $\G$ and between $H$ and $G$.

Linearized gauge transformations corresponding to the massless integer-superspin-$s$ superfields 
are given by the expressions
\begin{subequations} \label{4.2}
\bea
\d H_{\a(s-1)\ad(s-1)} &=& \cD^\b L_{\b \a(s-1) \ad(s-1)} - \cDB^\bd \bar{L}_{\a(s-1)\bd\ad(s-1)} \ ,\label{4.2a}\\
\d \G_{\b \a(s-1) \bd\ad(s-1)} &=& \frac{s+1}{2(s+2)} \cDB^\gd \cDB_{(\gd} \cD_{(\b} \bar{L}_{\a(s-1))\bd\ad(s-1) )} \label{4.2b}\\
&&+ \ri (s+1) \cDB^\gd \cD_{(\b(\gd} \bar{L}_{\a(s-1))\bd\ad(s-1))} \ , \non\\
\d G_{\b \a(s-1) \bd \ad(s-1)} &=& \hf \cDB_{(\bd} \cD_{(\b} \cD^{|\g|} L_{\a(s-1))\g\ad(s-1))} \ , \label{4.2c}
\eea
\end{subequations}
where $L_{\a(s)\ad(s-1)}$ is an arbitrary fermionic superfield of the Lorentz type $(s/2, (s-1)/2)$. 
The superscript $\g$ in \eqref{4.2c} is not subject to symmetrization. 
The guiding principle in choosing the form of transformations for $\G$ and $G$ is the requirement 
that the variations $\d \G$ and $\d G$ should be expressed in terms of the transverse linear superfield
\be 
\cDB^\bd \bar{L}_{\a(s-1)\bd\ad(s-1)} \non
\ee
and its conjugate, which arise in \eqref{4.2a}. 
This requirement, supplemented with the condition of transverse linearity,
 fixes the transformation \eqref{4.2b}, for it can be rewritten as
\bea \d \G_{\b \a(s-1) \bd\ad(s-1)} = - \hf \cD_{(\b} \cDB_{(\bd} \cDB^{|\gd|} \bar{L}_{\a(s-1))\ad(s-1) ) \gd} 
+ \ri s \cD_{(\b ( \bd} \cDB^{|\gd|} \bar{L}_{\a(s-1))\ad(s-1))\gd } \ , \non
\eea
where the superscript $\gd$ is not subject to symmetrization. The structure of the variation \eqref{4.2c} can be modified by making the substitution $G \rightarrow \widetilde{G}$, where 
\be 
\widetilde{G}_{\b\a(s-1)\bd\ad(s-1)} 
= G_{\b\a(s-1)\bd\ad(s-1)} + c\,  \cDB_{(\bd} \cD_{(\b} H_{\a(s-1)) \ad(s-1))}  \non
\ee
and $c$ is an arbitrary constant.

The quadratic functional of the superfields \eqref{4.1a} that is invariant under 
the transformations \eqref{4.2a} and \eqref{4.2b} has the form
\bea
S^{\perp}_{(s)} &=& - \Big( - \hf \Big)^s \int \rd^8z \,E^{-1} 
\Big\{ - \frac{1}{8} H^{\a(s-1) \ad(s-1)} \cD^\b (\cDB^2 - 4 \mu) \cD_\b H_{\a(s-1)\ad(s-1)} \non\\
&&+ \frac{1}{8} \frac{s^2}{(s+1)(2s+1)} [\cD^\b, \cDB^\bd] H^{\a(s-1)\ad(s-1)} [\cD_{(\b}, \cDB_{(\bd}] H_{\a(s-1)\ad(s-1))} \non\\
&&+\hf \frac{s^2}{s+1} \cD^{\b\bd} H^{\a(s-1)\ad(s-1)} \cD_{(\b(\bd} H_{\a(s-1))\ad(s-1))}
- \frac{(s+1)^2}{2} \mub\mu H \cdot H
 \non\\
&&+ \frac{2 \ri s}{2s+1} H^{\a(s-1)\ad(s-1)} \cD^{\b\bd} (\G_{\b\a(s-1)\bd\ad(s-1)} 
- \bar{\G}_{\b\a(s-1)\bd\ad(s-1)}) \non\\
&& 
+ \frac{2}{2s+1} \bar{\G} \cdot \G 
- \frac{s}{(s+1)(2s+1)} (\G \cdot \G + \bar{\G} \cdot \bar{\G}) \Big\} \ . 
\label{4.3}
\eea
The functional of the superfields \eqref{4.1b} that is invariant under 
the transformations \eqref{4.2a} and \eqref{4.2c} has a much simpler form, namely, 
\bea
S^{||}_{(s)} &=& \Big( - \hf \Big)^s \int \rd^8 z \,E^{-1} 
\Big\{ \frac{1}{8} H^{\a(s-1)\ad(s-1)} \cD^\b (\cDB^2 - 4 \mu) \cD_\b H_{\a(s-1)\ad(s-1)} \non\\
&&+ \frac{s}{s+1} H^{\a(s-1)\ad(s-1)} (\cD^\b \cDB^\bd G_{\b\a(s-1)\bd\ad(s-1)} - \cDB^\bd \cD^\b \bar{G}_{\b\a(s-1)\bd\ad(s-1)}) 
\non\\&&
+ \frac{(s+1)^2}{2} \mub \mu H \cdot H + 2 \bar{G} \cdot G + \frac{s}{s+1} (G \cdot G + \bar{G} \cdot \bar{G}) \Big\} \ . \label{4.4}
\eea
Both functionals generate equivalent dynamics because they are
related to each other by a duality transformation through the auxiliary model with action 
functional
\bea S[H, \G, V] &=& \left( - \hf \right)^s \int \rd^8 z E^{-1} \Big\{ \frac{1}{8} H^{\a(s-1)\ad(s-1)} \cD^\b (\cDB^2 - 4 \mu) \cD_\b H_{\a(s-1)\ad(s-1)} \non\\
&&+ \frac{(s+1)^2}{2} \mub \mu H \cdot H + \frac{1}{s+1} (s H^{\a(s-1)\ad(s-1)} \cD^\b \cDB^\bd V_{\b\a(s-1)\bd\ad(s-1)} \non\\
&&+ 2 \G \cdot V + (s+1) \bar{V} \cdot V +s V \cdot V + \rm{h.c.}) \Big\} \ , \label{4.5}
\eea
where $V_{\a(s)\ad(s)}$ is a complex unconstrained superfield of the Lorentz type $(s/2,s/2)$.

In the limit of vanishing curvature, the dynamical variables \eqref{4.1}, 
the gauge transformations \eqref{4.2}, and the gauge invariant action functionals 
\eqref{4.3} and \eqref{4.4} reduce to their flat-space analogs determining 
the two formulations \cite{KS} of the theory of a free massless super-spin-$s$ 
multiplet in the flat superspace. The component analysis shows that, 
after elimination of the auxiliary fields, the action functionals \eqref{4.3} and \eqref{4.4} 
describe free massless fields with spin $s$ and $s+1/2$ in the AdS space. 
Thereby, the functionals \eqref{4.3} and \eqref{4.4} realize two Lagrangian 
formulations of the theory of the superspin-$s$ multiplet. 
The formulation of the theory in terms of the superfields \eqref{4.1a} and the action functional 
\eqref{4.3} and that using the superfields \eqref{4.1b} and the action functional \eqref{4.4} 
can naturally be referred to as the transverse and longitudinal 
formulations, respectively. In both formulations, the gauge transformations have 
linearly dependent generators of an infinite stage of reducibility.

The physical sector has the simplest realization in the longitudinal formulation. 
It is described by two gauge-invariant field strengths, namely, the chiral 
tensor superfield
\bea
W_{\a_1 \cdots \a_{2s}} &=& \frac{\ri s}{2} \cD_{(\a_1}{}^{\bd_1} \cdots \cD_{\a_{s-1}}{}^{\bd_{s-1}} \cDB^{\bd_s} \cD_{\a_s} G_{\a_{s+1}\cdots \a_{2s}) \bd_1 \cdots \bd_s} \non\\
&&+\cD_{(\a_1}{}^{\bd_1} \cdots \cD_{\a_s}{}^{\bd_s} G_{\a_{s+1} \cdots \a_{2s}) \bd_1 \cdots \bd_s} \label{4.6}
\eea
and its conjugate. The explicit form of the superfield strength $W_{\a(2s)}$ in the transverse formulation 
can be found using the duality transformation.

\section{Component analysis}
In this section, we will determine the component structure of the theory described by the action functional \eqref{3.4}. The component analysis of the three other theories 
described by the functionals \eqref{3.6}, \eqref{4.3} and \eqref{4.4} can be carried out using the same method.

The most convenient way to recast a certain superfield action functional $S = \int \rd^8z E^{-1} \cL$ into the component form is to use Wess-Zumino gauge \eqref{2.6} for 
the background geometry of the AdS superspace. 
In this gauge, superspace and space-time covariant derivatives are related by equation \eqref{2.9}, 
and the component Lagrangian $L$ has the form (see \cite{Ideas} for more details)
\bea
\int\rd^8z \,E^{-1} \cL &=& \int\rd^8z \,\frac{E^{-1}}{\mu} \cL_c 
= -\frac{1}{4} \int \rd^4x\,e^{-1} 
(\cD^2 -12\mub )\cL_c| \equiv \int\rd^4x \, e^{-1} L \ , \non\\
\cL_c &=& - \frac{1}{4} (\cDB^2 - 4 \mu) \cL \ . \label{5.1}
\eea

In the theory described by the action functional \eqref{3.4}, 
the gauge freedom \eqref{3.2a} and \eqref{3.2b} 
can be used to algebraically  gauge away some of the 
component fields of $H$, $\G$, and $\bar{\G}$. 
Namely, we can choose the Wess-Zumino gauge 
in which only the fields that are given below and their conjugates  do not vanish:
\bea
b_{\b\a(s)\bd\ad(s)} &=& \hf [\cD_{(\b}, \cDB_{(\bd}] H_{\a(s)) \ad(s))}| = \bar{b}_{\b\a(s)\bd\ad(s)} \ , \non\\
\c_{\b\a(s)\ad(s)} &=& - \frac{1}{4} \cDB^2 \cD_{(\b} H_{\a(s) ) \ad(s)}| \ , \non\\
A_{\a(s)\ad(s)} &=& \frac{1}{32} \{ \cD^2, \cDB^2 \} H_{\a(s) \ad(s)}| = \bar{A}_{\a(s)\ad(s)} \ , \non\\
h_{\a(s-1)\ad(s-1)} &=& \G_{\a(s-1)\ad(s-1)}| = \bar{h}_{\a(s-1)\ad(s-1)} \ , \non\\
\bar{\vf}_{\b\a(s-1)\ad(s-1)} &=& \cD_{(\b} \G_{\a(s-1))\ad(s-1)}| \ , \non\\
\bar{\psi}_{\a(s-2)\ad(s-1)} &=& - \frac{s-1}{s} \cD^\g \G_{\g\a(s-2)\ad(s-1)}| \ , \non\\
\bar{\l}_{\a(s-1)\bd\ad(s-1)} &=& - \cDB_\bd \G_{\a(s-1)\ad(s-1)}| \ , \non\\
B_{\a(s-1)\ad(s-1)} &=& - \frac{1}{4} \cD^2 \G_{\a(s-1)\ad(s-1)}| \ , \non\\
F_{\b\a(s-1)\bd\ad(s-1)} &=& \cD_{(\b} \cDB_{\bd} \G_{\a(s-1))\ad(s-1)}| \ , \non\\
G_{\a(s-2)\bd\ad(s-1)} &=& \frac{s-1}{s} \cD^\g \cDB_\bd \G_{\g\a(s-2)\ad(s-1)}| \ , \non\\
\bar{\r}_{\a(s-1)\bd\ad(s-1)} &=& \frac{1}{4} \cD^2 \cDB_\bd \G_{\a(s-1)\ad(s-1)}| \ . \label{5.2}
\eea
All of these component fields are completely symmetric with respect to the undotted spinor indices, 
and separately, with respect to the dotted ones. The fields 
$\bar{\l}$, $F$, $G$, and $\bar{\r}$ are symmetric with respect to the dotted indices 
by virtue of transverse linearity of $\G$ \eqref{2.15a}. The component Lagrangian 
$L^{\perp}_{(s+1/2)}$ obtained from the action functional \eqref{3.4} decomposes 
into the sum of the Lagrangians of fermionic and bosonic component fields, 
\be
L^{\perp}_{(s+1/2)} = L_f(\c, \vf, \psi, \r, \l) + L_b(b, h, A, B, F, G) \ .
\ee

The Lagrangian of the fermionic fields has the form
\bea
L_f &=& \Big( - \hf \Big)^s \Bigg\{ \ri \, \bar{\c}^{\a(s)\bd\ad(s)} \nabla^\b{}_\bd \c_{\b\a(s)\ad(s)} 
+ \ri \, \vf^{\b\a(s-1)\ad(s-1)} \nabla_\b{}^\bd \vf_{\a(s-1)\bd\ad(s-1)} \non\\
&&-\ri \bar{\psi}^{\a(s-2) \bd \ad(s-2)} \nabla^\b{}_\bd \psi_{\b\a(s-2)\ad(s-2)} 
-\ri \bar{\l}^{\a(s-1)\bd\ad(s-1)} \nabla^\b{}_\bd \l_{\b\a(s-1)\ad(s-1)} \non\\
&&- \Big[\ri \nabla_{\b\bd} \bar{\vf}^{\b\a(s-1)\bd\ad(s-2)} \psi_{\a(s-1)\ad(s-2)} 
+ \ri \nabla_{\b\bd} \bar{\c}^{\b\a(s-1) \bd \ad(s)} \bar{\l}_{\a(s-1)\ad(s)} \non\\
&&+ \Big( \frac{s+1}{s} \l^{\a(s)\ad(s-1)} - \bar{\vf}^{\a(s)\ad(s-1)}\Big) \r_{\a(s)\ad(s-1)} \non\\
&&-\frac{\mub}{2} s \c^{\a(s+1)\ad(s)} \c_{\a(s+1)\ad(s)} + \frac{\mub}{2}(s+1) \vf^{\a(s-1)\ad(s)} \vf_{\a(s-1)\ad(s)} \non\\
&&+ \frac{\mub}{2} \frac{s(s+1)}{s-1} \psi^{\a(s-1)\ad(s-2)} \psi_{\a(s-1)\ad(s-2)} 
+ \frac{3}{2} \mu \frac{s+1}{s} \l^{\a(s)\ad(s-1)} \l_{\a(s)\ad(s-1)} \non \\
&&- \frac{\mub}{2} (s-3) \bar{\l}^{\a(s-1)\ad(s)} \vf_{\a(s-1)\ad(s)} + \rm{h.c.} \Big] \Bigg\} \ .
 \label{5.3}
\eea
We see that the fields $\l$ and $\r$ are auxiliary. The remaining fields
\bea
&&\c_{\a(s+1)\ad(s)} \ , \quad \vf_{\a(s-1)\ad(s)} \ , \quad \psi_{\a(s-1)\ad(s-2)} \ , \non\\
&&\bar{\c}_{\a(s)\ad(s+1)} \ , \quad \bar{\vf}_{\a(s)\ad(s-1)} \ , \quad \bar{\psi}_{\a(s-2)\ad(s-1)} \label{5.4}
\eea
form the set of dynamical variables that are necessary for formulating 
the theory of a massless field with the half-integer spin $s+1/2$ \cite{FF2}. After elimination of the 
auxiliary fields, the Lagrangian $L_f$ takes the form
\bea
L_f' &=& \Big( - \hf \Big)^s \Bigg\{ \ri \bar{\c}^{\a(s)\bd\ad(s)} \nabla^\b{}_\bd \c_{\b\a(s)\ad(s)} 
+ \ri \frac{2 s + 1}{(s+1)^2} \bar{\vf}^{\b\a(s-1)\ad(s-1)} \nabla_\b{}^\bd \vf_{\a(s-1)\bd\ad(s-1)} \non\\
&&- \ri \bar{\psi}^{\a(s-2)\bd\ad(s-2)} \nabla^\b{}_\bd \psi_{\b\a(s-2)\a(s-2)} \non\\
&&- \Big[ \frac{\ri s}{s+1} \nabla_{\b\bd} \bar{\c}^{\b\a(s-1)\bd\ad(s)} \vf_{\a(s-1)\ad(s)} 
+\ri \nabla_{\b\bd} \bar{\vf}^{\b\a(s-1)\bd\ad(s-2)} \psi_{\a(s-1) \ad(s-2)} \non\\
&&+\frac{\mub}{2} s \c^{\a(s+1)\ad(s)} \c_{\a(s+1)\ad(s)} 
+ \frac{\mub}{2} \frac{2s+1}{s+1} \vf^{\a(s-1)\ad(s)} \vf_{\a(s-1)\ad(s)} \non\\
&&+\frac{\mub}{2} \frac{s(s+1)}{s-1} \psi^{\a(s-1)\ad(s-2)} \psi_{\a(s-1)\ad(s-2)} + \rm{h.c.} \Big] \Bigg\} \ . 
\label{5.5}
\eea
Up to a scale transformation of the fields \eqref{5.4}, this coincides 
with the Lagrangian of a massless field with the spin $s+1/2$ on the AdS space \cite{FF2}.

The Lagrangian of the  bosonic fields has the form
\bea
L_b &=& \Big( - \hf \Big)^s \Bigg\{ - \frac{1}{4} b^{\a(s+1) \ad(s+1)} \Box b_{\a(s+1)\ad(s+1)} \non\\
&&+ \frac{1}{8} \nabla_{\b\bd} b^{\b\a(s)\bd\ad(s)} \nabla^{\g\gd} b_{\g\a(s)\gd\ad(s)} 
+ \frac{s^2 - 3}{4} \mub\mu b\cdot b \non\\
&&+2 h^{\a(s-1)\ad(s-1)} \Box h_{\a(s-1)\ad(s-1)} + \hf \mub \mu \big(s^2 - 11 (2s+1)\big) h \cdot h \non\\
&&+ 2 A\cdot A + A \cdot (F+\bar{F}) \non\\
&&+ \frac{\ri}{4} \nabla_{\b\bd} b^{\b\a(s)\bd\ad(s)} (F_{\a(s)\ad(s)} - \bar{F}_{\a(s)\ad(s)}) + \hf \bar{F} \cdot F \non\\
&&+ 2 \bar{B} \cdot B + \Big[\frac{s+1}{4s} F\cdot F + \frac{s+1}{4(s-1)} G\cdot G \non\\
&&-\ri F^{\b\a(s-1)\bd\ad(s-1)} \nabla_{\b\bd} h_{\a(s-1)\ad(s-1)} - \mu (3 s + 1) h \cdot B \non\\
&&+ \ri h^{\a(s-1)\bd\ad(s-2)} \nabla^\b{}_\bd \bar{G}_{\b\a(s-1)\ad(s-2)} + \rm{h.c.} \Big] \Bigg\} \ ,
 \label{5.6}
\eea
where $\Box = \nabla^a \nabla_a$. It is seen from this expression that the component fields 
$A$, $B$, $F$, and $G$ are auxiliary fields. The remaining real spin-tensors 
\be b_{\a(s+1)\ad(s+1)} \ , \quad h_{\a(s-1)\ad(s-1)} \label{5.7}
\ee
form the set of dynamical variables that are necessary for formulating 
the theory of a massless field with the spin $s+1$ \cite{Fronsdal3}. 
Eliminating the auxiliary fields from \eqref{5.6}, we obtain
\bea
L_b' &=& \Big( - \hf \Big)^s \Bigg\{ - \frac{1}{4} b^{\a(s+1) \ad(s+1)} \Box b_{\a(s+1)\ad(s+1)} \non\\
&&+ \frac{s+1}{8} \nabla_{\b\bd} b^{\b\a(s)\bd\ad(s)} \nabla^{\g\gd} b_{\g\a(s)\gd\ad(s)} \non\\
&&+s b^{\b\g\a(s-1)\bd\gd\ad(s-1)} \nabla_{\b\bd}\nabla_{\g\gd} h_{\a(s-1)\ad(s-1)} \non\\
&&+4\frac{2s+1}{s+1} h^{\a(s-1)\ad(s-1)} \Box h_{\a(s-1)\ad(s-1)} \non\\
&&+2 \frac{(s-1)^2}{s+1} \nabla_{\b\bd} h^{\b\a(s-2)\bd\ad(s-2)} \nabla^{\g\gd} b_{\g\a(s-2)\gd\ad(s-2)} \non\\
&&+\frac{s^2-3}{4} \mub \mu b\cdot b - 4(s+1)(2s+1)\mub\mu h\cdot h  \Bigg\} \ . \label{5.8}
\eea
Up to a scale transformation of the fields \eqref{5.7}, this expression coincides with the Lagrangian of 
a massless field with the spin $s+1$ on the AdS space 
\cite{Fronsdal3}.

\section{Conclusion}

In this paper, we constructed two dually equivalent formulations of the theory of free massless 
half-integer-superspin $(s + 1/2, s \geq 2)$ and integer-superspin 
$(s, s \geq 1)$ multiplets in AdS superspace. Both formulations from Section 3 
are well suited for describing the massless superspin $3/2$ 
(which corresponds to linearized supergravity), although this case $(s = 1)$ is rather peculiar. 
Let us first discuss the situation taking place in the case of 
the longitudinal formulation. Here, a specific feature is that, for $r = t = s - 1$ and $s = 1$, 
equation \eqref{2.15b} means chirality, i.e., $\cDB_\bd G = 0$. 
Furthermore, it is easy to verify that the gauge transformation \eqref{3.2c} can be recast 
into the equivalent form
\bea
\d G_{\a(s-1) \bd \ad(s-2)} = &-& \frac{1}{4} (\cDB^2 - 2 (s+1) \mu) 
\cD^\g L_{\g \a(s-1) \bd \ad(s-2)} \non\\
&+& \ri (s-1) \cDB_{(\bd} \cD^{\g| \gd|} L_{\g \a(s-1) \ad(s-2) ) \gd} \ . \non
\eea
It follows that at $s = 1$ the expression on the right-hand side becomes chiral. 
Thus, in the case of $s = 1$, the longitudinal formulation for half-integer superspins is 
specified by the dynamical variables
\be v^{||}_{3/2} = \left\{ H_{\a\ad}(z), G(z), \bar{G}(z) \right\} \ , \quad \cDB_\ad G = 0 \ ,
\ee
which are gauge superfields with respect to the transformations
\bea
\d H_{\a\ad} &=& \cDB_\ad L_\a - \cD_\a \bar{L}_\ad \ , \non\\
\d G &=& - \frac{1}{4} (\cDB^2 - 4 \mu) \cD^\b L_\b \ ,
\eea
where $L_\a$ is an arbitrary spinor superfield. 
The corresponding gauge-invariant action is a special case of the action \eqref{3.6} and has the form
\bea S^{||}_{(3/2)} =&& \int \rd^8z \,E^{-1} 
\Big\{ -\frac{1}{16} H^{\a\ad} \cD^\b (\cDB^2 - 4 \mu) \cD_\b H_{\a\ad} \non\\
&&+\frac{1}{48} ([\cD_\a, \cDB_\ad] H^{\a\ad})^2 - \frac{1}{4} (\cD_{\a\ad} H^{\a\ad})^2 
- \frac{|\mu|^2}{4} H^{\a\ad} H_{\a\ad} \non \\
&&+ \frac{\ri}{3} \cD_{\a\ad} H^{\a\ad} (\bar{G} - G) 
-\frac{1}{3} (\bar{G} G - G^2 - \bar{G}^2) \Big\} \ .
\eea
It can be shown that $S^{||}_{(3/2)}$ is obtained by linearizing the action
\be S_{SG} = \frac{1}{\k^2} \int \rd^8z {\bm E}^{-1} \left\{ -3 + \frac{\mu}{2 R} + \frac{\mub}{2 R} \right\}
\ee
of minimal $N = 1$ supergravity with a cosmological term \cite{SG} with respect to the background superspace specified by the conditions \eqref{2.3}. 
Here $\k$ is the gravitational constant, $\bm E = \rm{Ber}(\bm E_A{}^M)$, 
where $\bm E_A{}^M$ is the supervierbein of minimal $N =1$ supergravity, and $R$ 
is one of the components of the corresponding supertorsion.

Let us now discuss the transverse formulation of the theory of half-integer superspins at $s = 1$. 
Here, a specific feature is that the constraint \eqref{2.15a} is 
not defined at $r = t = s - 1$ when $s = 1$. However, its corollary \eqref{2.17a} 
and the general solution \eqref{2.18a} are well defined as $s = 1$ and are 
consistent with the gauge transformation \eqref{3.2b}. Thus, at $s = 1$, 
the transverse formulation of the theory of half-integer spins is given in terms of the gauge 
superfields
\be v^{\perp}_{(3/2)} = \left\{ H_{\a\ad}(z), \G(z), \bar{\G}(z) \right\} \ , \quad (\cDB^2 - 4 \mu) \G = 0
\ee
obeying the gauge transformations
\be \d H_{\a\ad} = \cDB_\ad L_\a - \cD_\a \bar{L}_\ad \ , \quad \d \G = - \frac{1}{4} \cDB^\bd \cD^2 \bar{L}_\bd \ .
\ee
In the limit of vanishing curvature, the corresponding gauge-invariant action $S^{\perp}_{(3/2)}$ 
reduces to the linearized action of non-minimal $n = -1$  supergravity (see \cite{Ideas}).

The case of $s = 1$ also reveals certain specific features for the formulations describing integer-superspin multiplets, namely, at $s = 1$, and only in this case, 
the superfield $H$ can be completely gauged away using transformations \eqref{4.2a}. \\

\noindent
{\bf Acknowledgements:}\\
This work was carried out under financial support of the Russian Foundation for Fundamental Research, the 
International Science Foundation (grant no. M2I000), and the European Community (grant no. INTAS-93-2058).

\appendix

\section{Appendix}
\setcounter{equation}{0}
\numberwithin{equation}{section}

We follow the notation and conventions adopted in \cite{WB}. 
In particular, lower-case Latin letters are used to denote 
vector indices, and lower-case Greek letters denote spinor indices. 
Tangent (Lorentz) indices are denoted by letters 
from the beginning of the alphabets, and the world indices are denoted by the letters from their middle parts. 
The vector index of the covariant derivatives $\cD_a$ and $\nabla_a$ is often 
converted to a pair of spinor  indices as follows:
\be 
\cD_{\a\ad} = (\s^a)_{\a\ad} \cD_a \ , \qquad \cD_a = - \hf (\s_a)^{\ad\a} \cD_{\a\ad} \ .
\ee
The contractions of  two spinor covariant derivatives are defined according to the rule
\be \cD^2 = \cD^\a \cD_\a \ , \qquad \bar{\cD}^2 = \bar{\cD}_\ad \bar{\cD}^\ad \ .
\ee
The action of the generators $M_{\a\b}$ and $\bar{M}_{\ad\bd}$ of the Lorentz group 
on spinors is given by the relations
\bea
M_{\a\b} \psi_\g &= \hf (\ve_{\g\a} \psi_\b + \ve_{\g\b} \psi_\a) \ , \qquad
\bar{M}_{\ad\bd} \psi_\gd = \hf (\ve_{\gd \ad} \psi_{\bd} + \ve_{\gd\bd} \psi_\ad) \ .
\eea
The generators $M_{ab}$ with vector indices are related to $M_{\a\b}$ and $\bar{M}_{\ad\bd}$ by the equation
\bea M_{ab} = &-& M_{ba} = \hf (\s_a)_{\a\gd} (\tilde{\s}_b)^{\gd\b} M^\a{}_\b 
- \hf (\tilde{\s}_a)^{\ad\g} (\s_b)_{\g \bd} \bar{M}_\ad{}^\bd \ .
\eea

All tensor (super)fields encountered in this paper are completely symmetric 
with respect to their undotted and spinor indices, and separately, 
with respect to their dotted indices. We use the notation
\bea
V_{\a(r)\ad(t)} &=& V_{\a_1 \cdots \a_r \ad_1 \cdots \ad_t} = V_{(\a_1 \cdots \a_r)(\ad_1 \cdots \ad_t)} \ ,
\\
V \cdot U &=& V^{\a(r) \ad(t)} U_{\a(r)\ad(t)} = V^{\a_1 \cdots \a_r \ad_1 \cdots \ad_t} U_{\a_1 \cdots \a_r \ad_1 \cdots \ad_t} \label{A.6} \ , \\
V_{(\a(r)} U_{\b(s) )} &=& V_{(\a_1 \cdots \a_r} U_{\b_1 \cdots \b_s )} \ .
\eea
Parentheses denote symmetrization of indices; 
the undotted and dotted spinor indices are symmetrized independently. 
Indices sandwiched between vertical  bars (e.g. $|\gd|$) are not subject to symmetrization. 
Throughout the entire paper, 
we assume that (super)fields carrying an even number of spinor indices correspond to bosons, 
whereas (super)fields carrying an odd number of spinor indices correspond to fermions.

In our calculations, we use the (anti)commutation relations \eqref{2.2} 
and a number of identities following from them. These identities have the form
\bea
\cD_\a \cD_\b &=& \hf \ve_{\a\b} \cD^2 - 2 \mub M_{\a\b} \ , \non\\
\cDB_\ad \cDB_\bd &=& - \hf \ve_{\ad\bd} \cDB^2 + 2 \mu \bar{M}_{\ad\bd} \ , \non\\
\cDB_\ad \cDB^2 &=& 4 \mu \cDB^\bd \bar{M}_{\ad\bd} + 4 \mu \cDB_\ad \ ,  \non\\
\cDB^2 \cDB_\ad &=& - 4 \mu \cDB^\bd \bar{M}_{\ad\bd} - 2 \mu \cDB_\ad \ ;
\eea
\bea
\left[ \cDB^2, \cD_\a \right] &=& 4 \ri \cD_{\a\bd} \cDB^\bd + 4 \mu \cD_\a 
= 4 \ri \cDB^\bd \cD_{\a\bd} - 4 \mu \cD_\a \ , \non\\
\left[ \cD^2, \cDB_\ad \right] &=& - 4 \ri \cD_{\b \ad} \cD^\b + 4 \mub \cDB_\ad 
= - 4 i \cD^\b \cD_{\b \ad} - 4 \mub \cDB_{\ad} \ ;
\eea
\bea
\cD^\a (\cDB^2- 4 \mu) \cD_\a = \cDB_\ad (\cD^2 - 4 \mub) \cDB^\ad \ .
\eea
In particular, the last identity guaranties the fulfillment of the requirement that the superfield action functionals constructed in Sections 3 and 4 
should be real-valued.

In a special gauge (with respect to \eqref{2.5}), the spinor vierbeins $E_\a$ and $\bar{E}^\ad$ 
corresponding to  the covariant derivatives $\cD_\a$ 
and $\cDB^\ad$ \eqref{2.1} are expressed in terms of the gravitational superfield $\cH^m(z)$ and the chiral compensator $\vf(z) = \vf(x + \ri \cH, \q)$  in terms of the semi-covariant supervierbein
\bea
\hat{E}_A &=& (\hat{E}_a, \hat{E}_\a, \hat{\bar{E}}^\ad) \ , \non\\
\hat{E}_\a &=& \d^\mu_\a \{ \partial_\mu + \ri (\partial_\mu \cH^n) (1- \ri \partial \cH)^{-1}{}_n{}^m \partial_m \} \ , \non\\
\hat{\bar{E}}^\ad &=& \d^\ad_{\dot{\mu}} \{ \bar{\partial}^{\dot{\mu}} - \ri (\bar{\partial}^{\dot{\mu}} \cH^n)) (1 + \ri \partial \cH)^{-1}{}_n{}^m \partial_m) \} \ , \non\\
\hat{E}_a &=& - \frac{1}{4} (\tilde{\s}_a)^{\ad \a} \{ \hat{E}_\a, \hat{\bar{E}}_\ad \} = \hat{E}_a{}^m \partial_m \ ,
 \label{A.11}
\eea
where
\be
(1 \pm \ri \partial \cH)_n{}^m = \d^m_n \pm \ri \frac{\partial \cH^m}{\partial x^n} \ , \non
\ee
and the density
\bea
F = &\vf&^{1/2} \bar{\vf}^{-1} \det{}^{-1/3}(1- \ri \partial \cH) 
\det{}^{1/6}(1 + \ri \partial \cH) \det{}^{-1/6} (\hat{E}_a{}^m) \ . \label{A.12}
\eea
In the explicit form, we have
\be E_\a = F \hat{E}_\a \ , \qquad \bar{E}^\ad = \bar{F}\hat{\bar{E}}^\ad \ . \label{A.13}
\ee
The connections $\Omega_\a$ and $\bar{\Omega}^\ad$ are determined from the (anti)commutation relations \eqref{2.2}. In the gauge $\vf = 1$, the expressions 
\eqref{A.11} -- \eqref{A.13}  reduce to the Ogievetsky-Sokatchev structure blocks \cite{OS2}.

In the gauge specified by the relations \eqref{2.11} and \eqref{2.12}, 
the explicit form of the generators of the superalgebra  $\rm osp(1, 4)$ is given by the expressions
\bea
{\bm P}_{\a\ad} = &&- \ri \partial_{\a\ad} - \frac{\ri}{8} \mub\mu \left(x_{(+)}{}_\a{}^\bd x_{(+)}{}^\b{}_\ad + x_{(-)}{}_\a{}^\bd x_{(-)}{}^\b{}_\ad\right) \partial_{\b \bd} \non\\
&&+ \frac{\ri}{2} \mub \mu \left(\q_\a x_{(+)}{}^\b{}_\ad \partial_\b + \bar{\q}_\ad x_{(-)}{}_\a{}^\bd \bar{\partial}_\bd\right) \non\\
&&- \frac{\ri}{2} \mub \mu \left(x_{(+)}{}^\b{}_\ad M_{\a\b} + x_{(-)}{}_\a{}^\bd \bar{M}_{\ad\bd}\right) \ , \label{A.14} \\
{\bm J}_{\a\b} = &&- \frac{\ri}{4} \left(x_\a{}^\ad \partial_{\b \ad} + x_\b{}^\ad \partial_{\a\ad}) + \frac{\ri}{2} (\q_\a \partial_\b + \q_\b \partial_\a\right) - \ri M_{\a\b} \ , \non\\
{\bm Q}_\a = && \ri \left(1 - \bar{\mu} \q^2\right) \partial_\a + \qb^\ad \partial_{\a\ad} + \frac{\ri}{2} \mub x_{(+)}{}_\a{}^\bd \q^\b \partial_{\b\bd} \non\\
&&- \hf \mub x_{(-)}{}_\a{}^\bd \bar{\partial}_\bd + 2 \ri \bar{\mu} \q^\b M_{\a\b} \ . \non
\eea
Here $x^a_{(\pm)} = x^a \pm \ri \q \s^a \bar{\q}$, and all the partial derivatives are the flat-space ones. 
The generators $\bar{\bm J}_{\ad\bd}$ and $\bar{\bm Q}_\ad$ are obtained 
from ${\bm J}_{\a\b}$ and ${\bm Q}_\a$, respectively, by means of conjugation.

\begin{footnotesize}

\end{footnotesize}

\end{document}